\documentclass[a4paper,conference]{IEEEtran}
\usepackage{ifpdf}

\usepackage{cite}

\ifCLASSINFOpdf
   \usepackage[pdftex]{graphicx}
\else
\fi
 \usepackage{graphicx}
\usepackage{epstopdf}

\usepackage[cmex10]{amsmath}
\usepackage{algorithmic}

\usepackage{array}

\usepackage{mdwmath}
\usepackage{mdwtab}

\usepackage{eqparbox}

\usepackage[tight,footnotesize]{subfigure}

\usepackage[font=footnotesize]{subfig}
\usepackage{multirow}

\usepackage{fixltx2e}

\usepackage{stfloats}

\usepackage{url}

\begin{document}
%
\title{Direct Vehicle-to-Vehicle Communication with Infrastructure Assistance in 5G Network}

\author{\IEEEauthorblockN{Ji Lianghai, Man Liu, Andreas Weinand, Hans D. Schotten}
\IEEEauthorblockA{Chair of Wireless Communication, University of Kaiserslautern, Germany, $\lbrace$ji,manliu,weinand,schotten$\rbrace$@eit.uni-kl.de}}


%


\IEEEoverridecommandlockouts
\IEEEpubid{\makebox[\columnwidth]{\copyright~Copyright 2017
		IEEE \hfill} \hspace{\columnsep}\makebox[\columnwidth]{ }}

\maketitle

\begin{abstract}
Compared with today's 4G wireless communication network, the next generation of wireless system should be able to provide a wider range of services with different QoS requirements. One emerging new service is to exploit cooperative driving to actively avoid accidents and improve traffic efficiency. A key challenge for cooperative driving is on vehicle-to-vehicle (V2V) communication which requires a high reliability and a low end-to-end (E2E) latency. In order to meet these requirements, 5G should be evaluated by new key performance indicators (KPIs) rather than the conventional metric, as throughput in the legacy cellular networks. In this work, we exploit network controlled direct V2V communication for information exchange among vehicles. This communication process refers to packet transmission directly among vehicles without the involvement of network infrastructure in U-plane. In order to have a network architecture to enable direct V2V communication, the architecture of the 4G network is enhanced by deploying a new central entity with specific functionality for V2V communication. Moreover, a resource allocation scheme is also specifically designed to adapt to traffic model and service requirements of V2V communication. Last but not least, different technologies are considered and simulated in this work to improve the performance of direct V2V communication.
\end{abstract}


%
\IEEEpeerreviewmaketitle

\section{Introduction}
%
%
In Europe alone, around 40 000 people die and 1.7 million are injured annually in traffic accidents. At the same time, traffic increases on our roads leading to traffic jams, increased travel time, fuel consumption and increased pollution \cite{D11}. Cooperative intelligent traffic systems (C-ITS) can address these problems by warning drivers of dangerous situations and intervene through automatic braking or steering if the driver is unable to avoid an accident. Besides, cooperative driving applications, such as platooning (road-trains) and highly automated driving can reduce travel time, fuel consumption, and $\text{CO}_2$ emissions and also increase road safety and traffic efficiency. The C-ITS systems rely on timely and reliable exchange of information among vehicles.\\
In order to provide a wireless network to enable the information exchange process, the new generation of wireless network should be designed to offer a solution with a high degree of reliability and availability, in terms of data rate, latency or another Quality of Service (QoS) parameter \cite{D11}\cite{D211}. In order to meet the corresponding demand for new service types, such as vehicular communications, intensive research work has been performed to design the 5G network to fulfill requirements of ultra-high reliability communication (URC)\cite{D222}.\\  
Compared with conventional KPIs, e.g. overall cell throughput and per link data rate, which were representatively used to evaluate legacy wireless systems, V2V communication experience different technical challenges \cite{D62}. What is expected for V2V communication is 5 times reduced E2E latency with much higher reliability compared with the current 4G network.\\
As a critical technology for the 5G system, device-to-device (D2D) communication \cite{my1}\cite{my2} has been proposed to facilitate the V2V communication in 5G. In literature, some publications have already discussed the feasibility to provide V2V communication in 5G network \cite{3gpp}\cite{survey} and most of them looked at V2V communication from a more generic view.\\
In this work, we exploit direct V2V communication for information exchange among vehicles. In Sect.~\ref{SM}, details of system models used in our work are provided. Then we demonstrate our radio access network architecture and resource allocation scheme in Sect.~\ref{RAN}, which are specifically adapted to enable V2V communication. In order to improve system performance, some key technologies contributing to direct V2V communication are discussed in Sect.~\ref{RT}. Finally, numerical results of system level simulation are given in Sect.~\ref{NR} and conclusion of our work is drawn in Sect.~\ref{con}.
\section{System Model}\label{SM}
In order to improve traffic safety and efficiency, information of each vehicle should be collected by all vehicles in its proximity, e.g. constructions, road hazards. Thus, one efficient way for information exchange is to multicast the information of one vehicle. In this manner, all vehicles which are in proximity of the transmitter vehicle should listen to the multicasted packet simultaneously. This communication process corresponds to a point-to-multipoint communication where several receivers try to receive the same packet coming from one transmitter. In this section, we give a description of models used in this work.
\IEEEpubidadjcol
\subsection{Environmental model}
\begin{figure}[!t]
\centering
\includegraphics[width=3.4in]{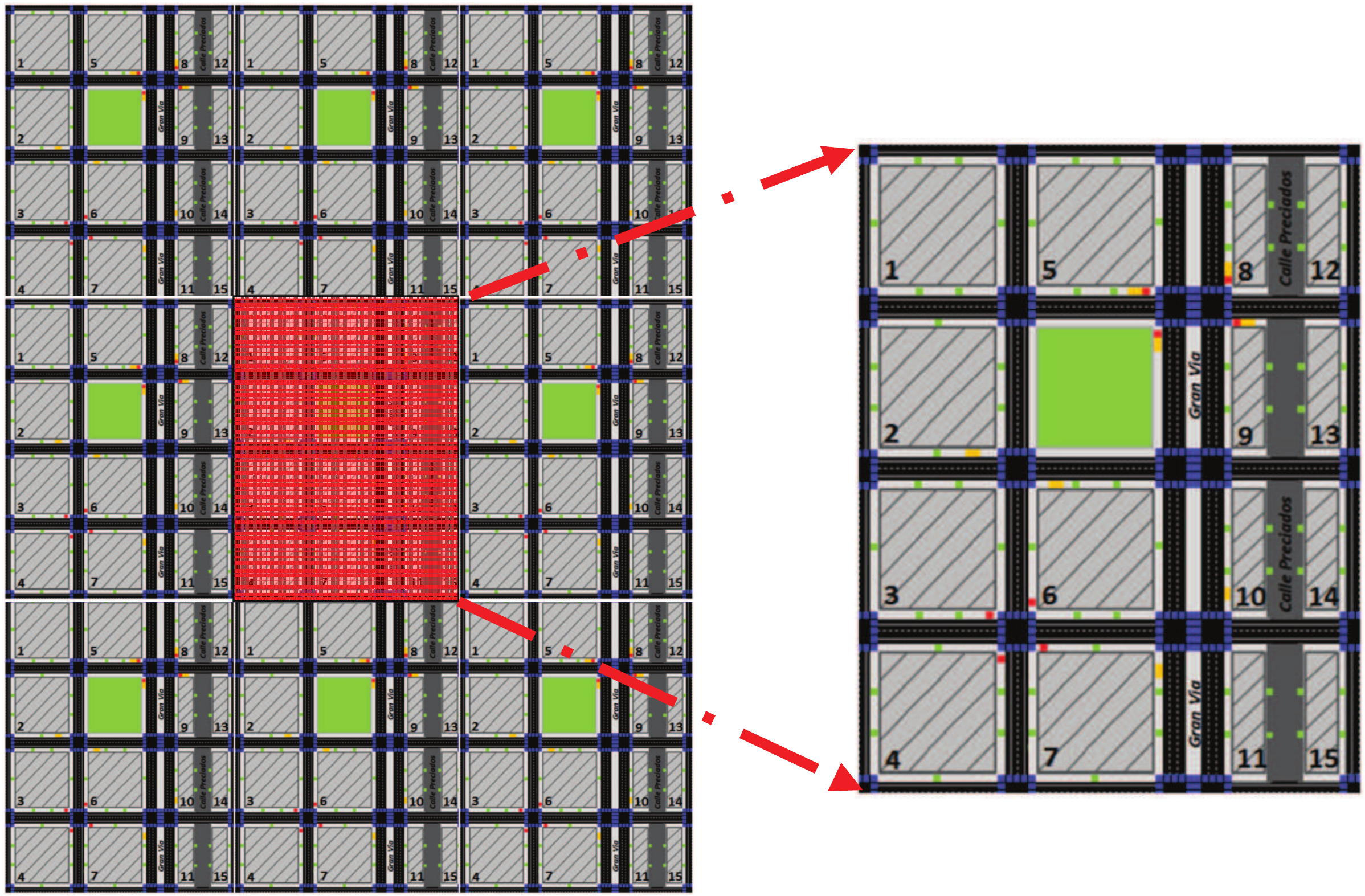}
\caption{Environment model}
\centering
\label{em}
\end{figure}

In this work, system performance of direct V2V communication is investigated in a dense urban environment. As shown in Fig.~\ref{em}, a Madrid grid environmental model defined in METIS project \cite{D61} is used here. In this model, a 3D visualization of Madrid grid is depicted where each grid composes of 15 buildings and one park. In order to avoid the cell border effect, another eight replicas of Madrid grid are also placed but only the performance of vehicles located in the central Madrid grid (as shown at right-hand side of Fig.~\ref{em}) are inspected to derive system performance from. Dimensions for one Madrid grid are 387 m (east-west) and 552 m (south-north). And building heights are uniformly distributed between 8 and 15 floors with 3.5 m per floor. A detailed description of this model can be found in \cite{D61}.
\subsection{Deployment model}
In Madrid grid, vehicles are distributed on road with a density of 1000 vehicles per square kilometer. An isotropic antenna is installed on each vehicle at 1.5-meter height. 1$\times$2 antennas configuration (receiver diversity) is exploited for V2V communication in this work and a target communication range of 200 meters is required. Each vehicle has a constant transmitting power of 24 dBm in each 10 MHz bandwidth. In this work, direct V2V communication operates on a carrier frequency of 5.9 GHz with a maximal bandwidth of 200 MHz. Due to a large number of retransmission trials experienced later in Sect.~\ref{NR}, where numerical results are shown, 100 Hz bandwidth is explicitly used for retransmission. Moreover, one Macro base station (BS) is deployed in each Madrid grid to provide control-plane (CP) functionalities (e.g., radio resource management). 
\subsection{Channel model}
In this work, a 3D channel propagation model in between vehicles is applied. Line-of-sight (LOS) propagation \cite{D61} and non-line-of-sight (NLOS) propagation \cite{NLOS} are both modeled in the Madrid grid.
\subsection{Traffic model and Mobility model}\label{TM}
Two traffic types are normally experienced in V2V communication. One is event-driven packet transmission, and the other one is periodic packet transmission. Since the event-driven type of traffic happens with a much lower frequency compared with the periodic transmission, it does not generate a lot of traffic. Therefore, a periodic packet transmission of 1600 Bytes with 10 Hz periodicity for each vehicle is considered in this work as traffic model \cite{D61}.\\
Due to the dimension and user density of Madrid grid, total number of vehicles in one Madrid grid is
\begin{align*}
N_V=1000\times (0.387\times 0.552)\approx 213~\text{vehicles}.
\end{align*}
and thus the overall date volume in one Madrid grid is approximately 27.3 Mbps.\\
Further, we assume that vehicles in this urban dense environment have a maximal velocity of 50 km/h. 
\section{Radio Access Network Architecture and Resource Allocation Scheme}\label{RAN}
Since V2V communication requires a very low latency with a very high reliability, legacy network architecture needs to be enhanced. Besides, due to the specific traffic for V2V communication, resource allocation scheme also needs to be adapted accordingly to improve the system performance. 
\subsection{Radio access network architecture}
\begin{figure}[!t]
\centering
\includegraphics[width=3.4in]{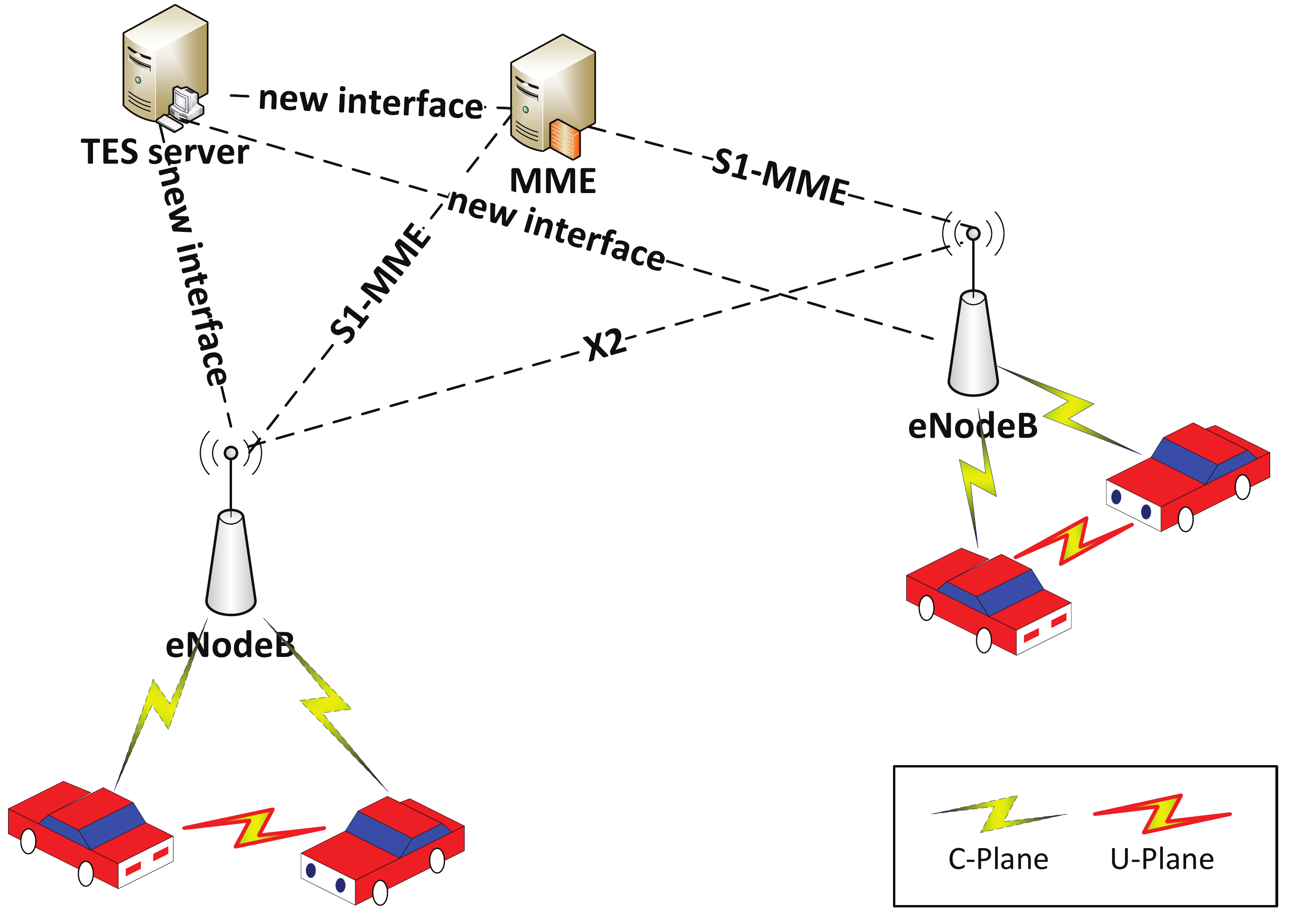}
\caption{System architecture}
\centering
\label{sa}
\end{figure}

In this work, network controlled direct V2V communication is exploited for packet transmission. As shown in Fig.~\ref{sa}, all vehicles are connected to operator network in CP. Meanwhile, user plane (UP) traffic communicates directly between the vehicle transmitter and the receivers who are located in the proximity of the transmitter. The direct V2V communication can contribute to a lower latency value since the network infrastructure is not involved. In order to provide network controlled direct V2V communication, we assume a traffic efficiency and safety (TES) server is installed in the core network of a mobile network operator and certain functionality (e.g., V2V control function) can be provided by this new server, as follows:
\begin{itemize}
\item{TES server can exchange location information with the mobility management entity (MME) which receives regularly location update messages from vehicles.} 
\item{Based on the location information of vehicles, this server can allocate resource to the BSs for V2V communication.}
\item{Based on real-time network environment and performance, TES server can adapt network settings for direct V2V communication.}
\end{itemize}
Due to the strict latency and reliability requirements of the considered V2V communication, the TES server needs to manage and adapt the network in a timely manner. Therefore, the front-haul connection between the TES server and the eNodeBs should be real-time and delay-critical.
\subsection{Resource allocation between first transmission and retransmission}
\begin{figure}[!t]
\centering
\includegraphics[width=3.4in]{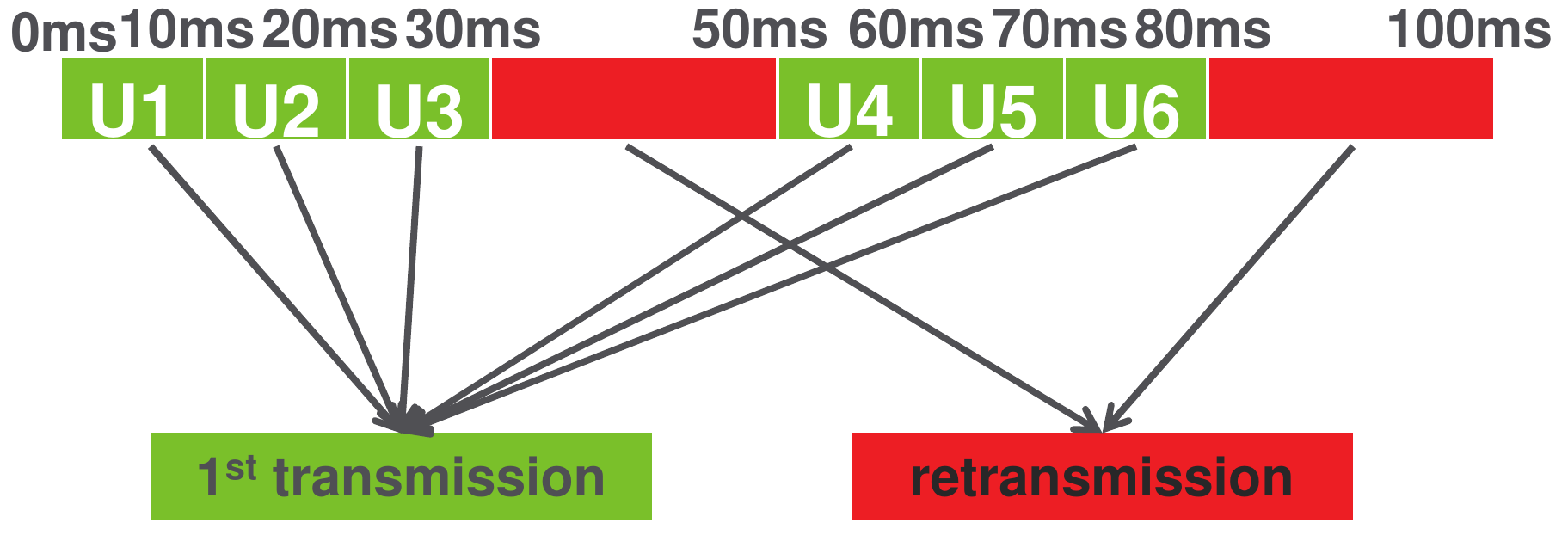}
\caption{Resource allocation between first transmission and retransmission}
\centering
\label{rrm}
\end{figure}

In order to support the V2V traffic with 10 Hz periodic transmission of a packet of 1600 Bytes, each packet transmission is assigned by the BS with certain resource. In this approach, the BS decides and pre-allocates resource to all vehicles. The basic principles for resource allocation are as follows: 
\begin{itemize}
\item{for first packet transmission: each transmitter is scheduled by the BS with a set of resource blocks (RBs), and its packet is generated in the transmission time interval (TTI) just before its transmission can start.}
\item{For retransmission: RBs are uniformly distributed among all RBs.}
\end{itemize}
One example is shown in Fig.~\ref{rrm} to express the above resource allocation procedure more clearly. In this example, six users (U1,U2,~$\ldots$~,U6) are assumed to be served by one BS in 100 ms (duration time of one period of 10 Hz) and we assume that first transmission of each packet occupies 10 ms of time resource, where in reality this time duration is related to packet length, coding and modulation rate, and also available transmission bandwidth. Meanwhile, if it is assumed that the retransmission requires a time duration of 20 ms, two blocks for retransmissions are available in this time duration of 100 ms and they are uniformly distributed. 
\section{Related Technologies}\label{RT}
Since the ultra-high reliability requirement of V2V communication poses novel challenges on the wireless network, some key technologies to enable direct V2V communication are discussed in this section.
\subsection{Modulation and coding scheme}
One issue for above point-to-multipoint communication is to adapt each transmission with an appropriate modulation and coding scheme (MCS). Due to the near-far effect, different links between one transmitter and its different receivers can experience different channel states. Therefore it is critical that links experiencing worse channel states should adapt to an MCS with a more robust link performance. Meanwhile, it brings a large signaling burden for the transmitter to collect the channel state information (CSI) of the receivers in the considered multicast communication scenario. Therefore, when the CSI is not available at the transmitter side, a more robust transmission means a lower modulation and coding rate. However, an MCS with lower rate requires larger resource to transmit one packet of a fixed length. Thus, the MCS should be dynamically adjusted by the network based on real-time system load. For instance, as we mentioned in Sect.~\ref{TM}, the date volume of 27.3 Mbps should be served in one Madrid grid with a bandwidth of 100 MHz for the first transmission. In this case, it requires a minimal spectral efficiency of 0.273 bits/Hz and an MCS with a spectral efficiency higher than this requirement should be selected by the network. More specifically, in the LTE network, it refers to the MCS with a spectral efficiency of 0.377 bits/Hz. Since this MCS scheme provides a low block error rate (BLER), i.e., lower than 10\%, to the links with signal-to-interference-plus-noise ratio (SINR) values higher than -3dB, retransmission will be required with high probability for the links with SINR values lower than -3dB.
\subsection{Context-aware coordination between the neighboring BSs}
\begin{figure}[!t]
\centering
\includegraphics[width=3.4in]{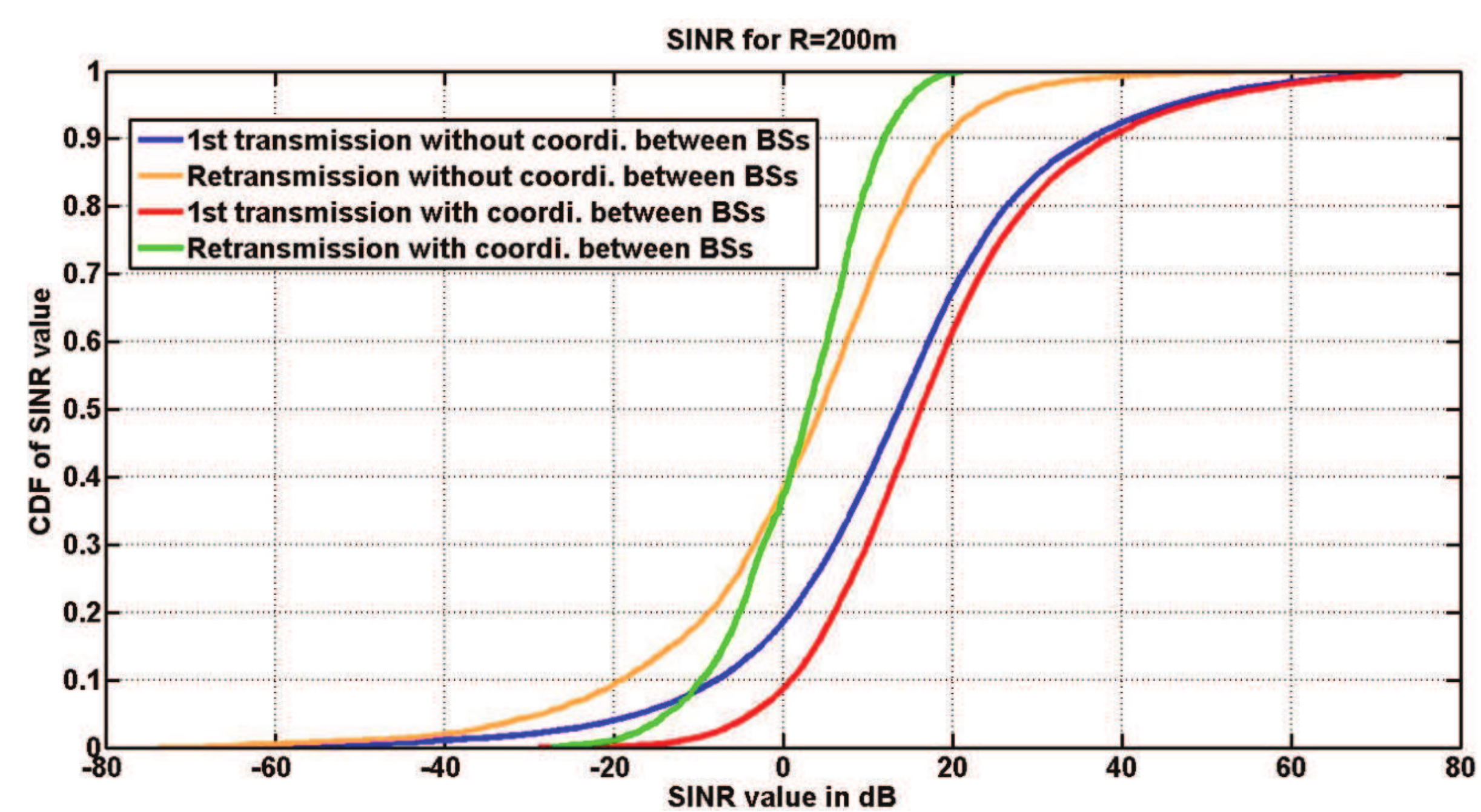}
\caption{System performance}
\centering
\label{sinr}
\end{figure}
\begin{figure}[!t]
\centering
\includegraphics[width=3.4in]{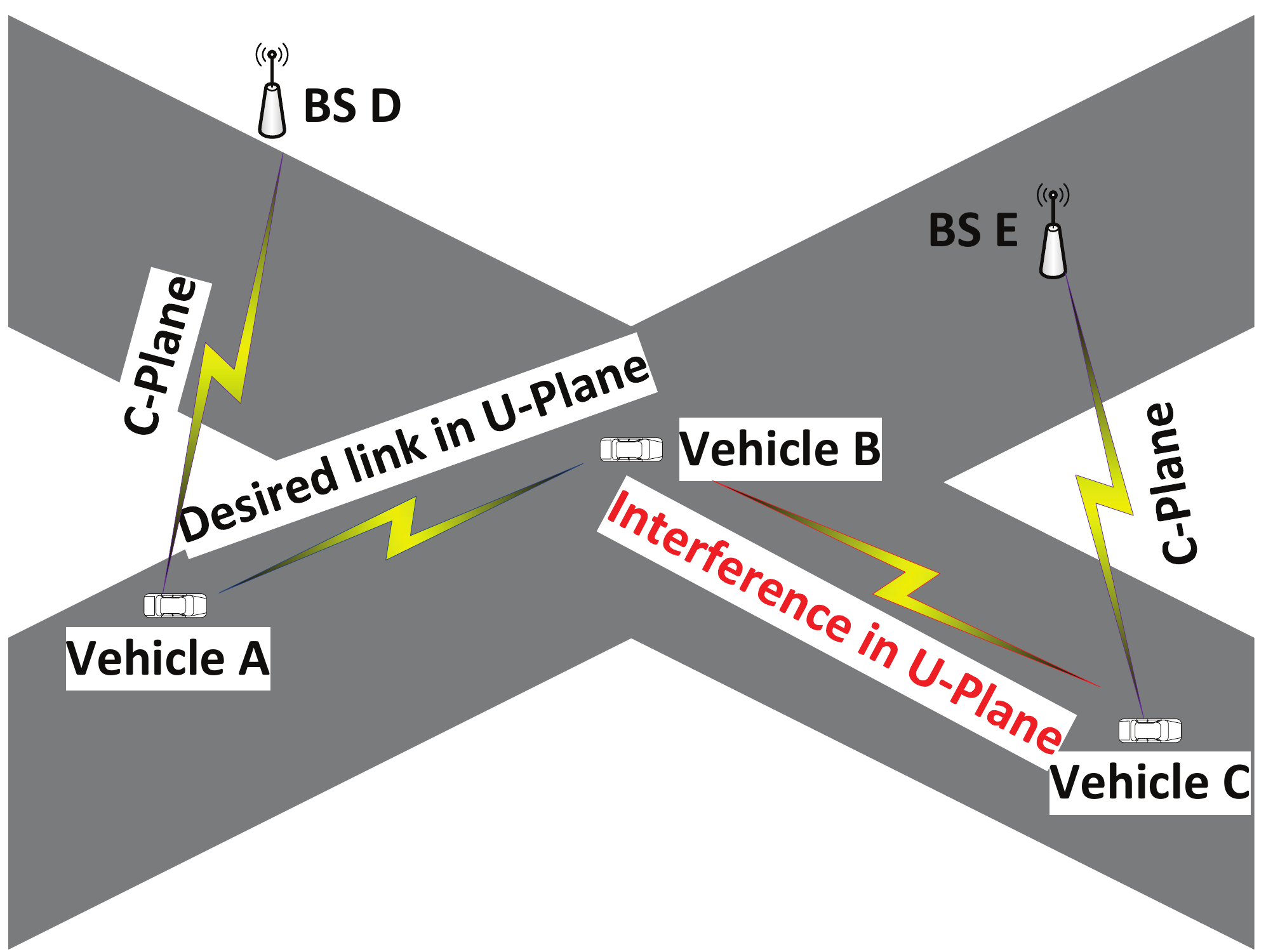}
\caption{Example to show disadvantages without coordination}
\centering
\label{PC}
\end{figure}
Now, we consider a scenario with a target V2V communication range up to 200 meters, and the cumulative distribution function (CDF) of the SINR values is plotted in Fig.~\ref{sinr}. It can be seen from this plot, in the case where no coordination between neighboring BSs is used, approximately 15 percent of receivers experience SINR values lower than -3 dB and they require retransmissions. To inspect on the performance of the retransmissions, their SINR values are also given in Fig.~\ref{sinr}. We can see that more than 30 percent of the retransmissions still experience SINR values lower than -3 dB. In this scenario, every BS has no knowledge of the situation in neighboring cells and thus two issues are presented:  
\begin{itemize}
\item {Without any awareness of interference coming from the neighboring cells, some receivers may experience strong interference.}
\item {When one receiver located on the border of one cell, it has a high probability to be in the proximity of another transmitter served by the neighboring cell, thus packet collision may occur.}
\end{itemize}
Fig.~\ref{PC} gives one example to show the above issues. Vehicle A and Vehicle C are two transmitters controlled by the different BSs D and E. Meanwhile, Vehicle B locates in the communication range of both Vehicles A and C. Without cooperation in between BS D and BS E, Vehicles A and C can transmit their packets simultaneously on the same time and frequency resource. Thus, these two packets collide at vehicle B and high interference can be experienced for each of the transmitted packets. To handle these issues, a context-aware coordination scheme between BSs can be exploited. The principle here is to allocate the same radio resource to the different transmitters only when these transmitters have an inter-transmitter distance $r$ larger than a threshold value. In this scheme, the context information refers to the vehicular positions. Moreover, in this work, the value of $r$ is set to be two times of the communication range $R$, in order to avoid packet collision. Fig.~\ref{sinr} also shows the SINR values of the proposed coordination scheme and it can be seen that around 9\% gain can be achieved by the coordination scheme, considering the successful ratio of the first transmission.
\subsection{Adaptive retransmission scheme}\label{rs}
Since the direct V2V communication corresponds to a point-to-multipoint transmission, the successful transmission ratio has a different impact on retransmission compared with the unicast transmission. The reason is that successful transmission ratio of $X$ in multicast does not mean that packets do not need to be retransmitted with a probability of $X$. For instance, one transmitter multicasts two packets to the nearby 10 vehicles where the two different vehicles fail in receiving the first and second packet transmission, respectively. Thus, a successful transmission ratio of 90\% is encountered here but both the two packets have to be retransmitted. Therefore, we can see that a larger number of packets need to be retransmitted in multicast mode compared with unicast mode, when the same successful transmission ratio is experienced.\\
Moreover, the successful transmission ratio decreases dramatically with an increased transmission distance due to a worse channel state. Therefore, a big portion of packets are required to be retransmitted, once the target transmission range achieves a certain level. In order to improve the link robustness for the retransmission, one MCS with lower modulation and coding rate (MCR) should be applied. However, more resource is required then, compared with the MCS with higher MCR. Therefore, a compromise in between the link robustness and available system resource should always be achieved by the V2V control function, taking the real-time system load into account.
\subsection{Shorted TTI length}\label{tti}
\begin{table}\caption{Delay component assumption}
\label{DC}
\begin{center}
\begin{tabular}{ |c|c| }
\hline Delay component &	Value in a unit of TTI \\\hline
frame alignment (FA)&	0.5 TTI\\\hline
\multirow{2}{*}{TTI per packet (TP)}&	Packet specific (depending on\\
& available spectrum resource)\\\hline
receiver processing delay (RPD)	& 1 TTI\\\hline
transmission of NACK feedback	&\multirow{2}{*}{1 TTI}\\
from receiver to BS (FB) & \\\hline
scheduling delay (SD) &	1 TTI\\\hline
transmission of retransmission &\multirow{2}{*}{1 TTI}\\
command from BS to Txs (TRC) &\\ \hline
\end{tabular}
\end{center}
\end{table}
In order to calculate E2E latency in U-plane, delay components shown in Tab.~\ref{DC} are considered in this work. All delay components are assumed with a unit of TTI and each TTI has a value of 1ms in the legacy LTE network. The minimal E2E latency happens when the first trial of packet transmission is successful and this packet transmission (TP) should cost only 1 TTI. Thus, a minimal E2E latency has a value of:
\begin{align*}
&\text{minimal E2E latency=0.5(FA)+1(TP)+1(RPD)=2.5 TTIs}
\end{align*}
In the case where the first packet transmission is failed, retransmission is required and the minimal E2E latency can be calculated as:
\begin{align*}
&\text{minimal E2E latency of retransmission=}\\
&\text{0.5(FA)+1(TP)+1(RPD)+1(FB)+1(SD)+1(TRC)+1(TP)+1(RPD)}\\
&\text{=7.5 TTIs}.
\end{align*}
Therefore, if each TTI has a duration of 1ms as in legacy LTE system, successfully retransmitted packets will have a larger value than 5ms, which exceeds the latency requirement of V2V communication \cite{D11}. One potential solution here is to decrease TTI duration from 1ms to a smaller value. Due to the better capability of the communication infrastructure deployed on vehicles, a certain level of TTI reduction is feasible in 5G.
\section{Numerical Results}\label{NR}
Tab.~\ref{results} demonstrates the system performance w.r.t. two different communication range requirements of 100 meters and 200 meters.  Successful packet transmission ratio and 5 ms end-to-end (E2E) latency ratio are used as the metrics to evaluate the system performance. Successful packet transmission ratio represents the probability of a packet successfully received at the receiver, no matter how much time it takes. While the 5 ms E2E latency ratio represents the probability of a packet successfully received with an E2E latency smaller than 5 ms. Moreover, the system parameters shown in Tab.~\ref{results} are as follows:
\begin{enumerate}
\item{no coordination between neighboring BSs,}
\item{coordination between neighboring BSs,}
\item{retransmission with spectral efficiency of 0.15 bits/Hz,}
\item{retransmission with spectral efficiency of 0.37 bits/Hz,}
\item{bandwidth=100 MHz,}
\item{bandwidth=200 MHz,}
\item{duration time of per TTI = 1 ms,}
\item{duration time of per TTI = 0.5 ms.} 
\end{enumerate}
As can be seen from this table, the system performance is more sensitive to the exploited technologies in the scenario with a larger communication range. Thus, we focus on the results where the communication range is up to 200 meters (R=200m) for analysis. In the first row, a system without any coordination between neighboring BSs with 100MHz bandwidth is exploited. Besides, modulation and coding scheme (MCS) used here for retransmission has a spectral efficiency of 0.15 bits/Hz and LTE frame structure is applied where per TTI has a time duration of 1 ms. It can be seen there, 85.59\% of the overall packets are received successfully within 5 ms E2E latency requirement and 86.26\% of overall packets are successfully received with any E2E latency values. In the second row, a coordination scheme is applied where location information of all vehicles are available in the central entity in order to mitigate interference and avoid packet collision. As can be seen here, the system performance has been improved compared with the previous case and 94.45\% of the packets can be successfully received within 5 ms. However, the successful ratio of 94.73\%, in this case, is not very much different from the value of 5ms E2E latency ratio. Actually, in the case where per TTI has a duration of 1ms, the difference between the two considered metrics represents the contribution from the successfully retransmitted packets. The reason why retransmission does not contribute much, in this case, is due to the fact that 100MHz bandwidth does not provide enough resource for the retransmissions and therefore a large ratio of the unsuccessfully received packets cannot be scheduled for retransmissions.\\
In row 3 and row 4, we increased system bandwidth to 200 MHz where 100 MHz is explicitly used for retransmission. Two different retransmission schemes are used where one has a spectral efficiency of 0.15 bits/Hz and the other of 0.37 bits/Hz. As can be seen here, since additional bandwidth is available for retransmissions, more retransmissions can be scheduled and therefore they contribute to a higher ratio of the overall successful transmission ratio. Moreover, by comparing the two cases shown in row 3 and row 4, we can see that though the MCS with a lower MCR provides a better link robustness compared with the other case, it has a successful ratio of 97.25\% which is lower than 98.5\% of the case where the MCS has a higher MCR. As mentioned in Sect.~\ref{rs}, the reason here is that the MCS with a higher MCR can schedule more retransmissions with the limited amount of resource. Besides, looking at the 5 ms E2E latency ratio, we can see that there is no change compared with the previous case (i.e., shown in the second row) where 100 MHz bandwidth is used. This demonstrates that the legacy LTE frame structure with a per TTI duration of 1 ms is not an optimal solution for direct V2V communication, since all the successfully retransmitted packets have E2E latency larger than 5 ms, as calculated in Sect~\ref{tti}.\\
In the last row, the TTI duration is decreased to from 1 ms to 0.5 ms. Due to this change, the retransmitted packets can have E2E latency lower than 5 ms and therefore contribute to the 5ms E2E latency ratio. Comparing with the previous cases, the 5ms E2E latency ratio can be improved from 94.45\% to 97.98\%.\\
It is to be noticed, though the above analysis is done w.r.t. the case where V2V communication range is up to 200 meters, it also applies to other cases where shorter communication range is required. 
\begin{table}\caption{System performance w.r.t. different schemes}
\centering
\label{results}
\begin{tabular}{ |c|c|c|c|c| }
\hline &	\multicolumn{2}{c|}{R=100m} & \multicolumn{2}{c|}{R=200m} \\\hline
\multirow{2}{*}{System} & \multirow{2}{*}{successful } & \multirow{2}{*}{5 ms E2E } & \multirow{2}{*}{successful } & \multirow{2}{*}{5 ms E2E } \\
parameters & ratio & ratio & ratio & ratio \\\hline
1),3),5),7) & 97.71\% & 96.82\% & 86.26\% & 85.59\% \\\hline
2),3),5),7) & 99.56\% & 99.18\% & 94.73\% & 94.45\% \\\hline
2),3),6),7) & 99.98\% & 99.18\% & 97.25\% & 94.45\% \\\hline
2),4),6),7) & 100\% & 99.18\%  & 98.5\% & 94.45\% \\\hline
2),4),6),8) & 100\% &99.98\% & 98.5\% & 97.98\% \\\hline
\end{tabular}
\end{table}
\section{Conclusion}\label{con}
In our work, we introduce the system architecture to enable the direct V2V communication under network control. Besides, a resource allocation scheme is also designed to dynamically adapt to the real-time traffic requirement of the V2V communication. Moreover, several key technologies are also proposed and evaluated to improve the system performance of the direct V2V communication. Last but not least, a system level simulator is built up and aligned with reality to produce reliable simulation results. Based on our evaluation work, it can be seen that all related technologies should add on top of each other to enable direct V2V communication and improve traffic safety and efficiency.\\


\section*{Acknowledgment}
A part of this work has been supported by the Federal Ministry of Education and Research of the Federal Republic of Germany (BMBF) in the framework of the project 5G-NetMobil. The authors would like to acknowledge the contributions of their colleagues, although the authors alone are responsible for the content of the paper which does not necessarily represent the project.




\begin{thebibliography}{1}
\bibitem{D11}
ICT-317669 METIS, Deliverable 1.1 Version 1, \emph{Scenarios, requirements and KPIs for 5G mobile and wireless system},  \hskip 1em plus 0.5em minus 0.4em\relax April 2013.
\bibitem{D211}
A. Osseiran et al., \emph{The Foundation of the Mobile and Wireless Communications System for 2020 and Beyond: Challenges, Enablers and Technology Solutions}, 2013 IEEE 77th Vehicular Technology Conference (VTC Spring), Dresden, 2013, pp. 1-5.
 \bibitem{D222}
METIS-II, Deliverable 2.1, \emph{Draft overall 5G RAN design},\hskip 1em plus 0.5em minus 0.4em\relax July 2016.
\bibitem{D62}
ICT-317669 METIS, Deliverable 6.2, \emph{Initial report on horizontal topics, first results and 5G system concept}, \hskip 1em plus 0.5em minus 0.4em\relax  March 2014.
\bibitem{my1}
Ji Lianghai, A. Klein, N. Kuruvatti, H. D. Schotten, \emph{System Capacity Optimization Algorithm for D2D Underlay Operation}, IEEE International Conference on Communications (ICC), Sydney, Australia, June 2014.
\bibitem{my2}
Ji Lianghai, Man Liu, H. D. Schotten, \emph{Context-aware Cluster Based Device-to-Device Communication to Serve Machine Type Communications}, IEEE International Conference on Communications (ICC), Paris, France, May 2017.
\bibitem{3gpp}
Vinel, A. \emph{3GPP LTE Versus IEEE 802.11p/WAVE: Which Technology is Able to Support Cooperative Vehicular Safety Applications} Wireless Communications Letters, IEEE , vol.1, no.2, pp.125,128, April 2012.
\bibitem{survey}
Araniti, G.; Campolo, C.; Condoluci, M.; Iera, A.; Molinaro, A. \emph{LTE for vehicular networking: a survey} Communications Magazine, IEEE , vol.51, no.5, pp.148,157, May 2013
\bibitem{D61}
ICT-317669 METIS, Deliverable 6.1 Version 1, \emph{Simulation guidelines},  \hskip 1em plus 0.5em minus 0.4em\relax October 2013.
\bibitem{NLOS}
Thomas Mangel, Oliver Klemp, Hannes Hartenstein. \emph{5.9 GHz inter-vehicle communication at intersections: a validated non-line-of-sight path-loss and fading model}. EURASIP Journal on Wireless Communications and Networking, 2011.
\end{thebibliography}
%

\end{document}